\documentclass[aps,epsf,twocolumn,showpacs]{revtex4}
\usepackage{amsmath}
\usepackage{epsfig}

\begin{document}

\title{Universality in disordered systems: The case of the three-dimensional random-bond Ising model}

\author{Nikolaos G. Fytas}\thanks{nfytas@phys.uoa.gr}\affiliation{Department of Materials Science,
University of Patras, Patras 26504, Greece}

\author{Panagiotis E. Theodorakis} \affiliation{Institut f\"{u}r Physik, Johannes Gutenberg-Universit\"{a}t, D-55099 Mainz, Germany}

\date{\today}

\begin{abstract}
We study the critical behavior of the $d=3$ Ising model with bond
randomness through extensive Monte Carlo simulations and
finite-size scaling techniques. Our results indicate that the
critical behavior of the random-bond model is governed by the same
universality class as the site- and bond-diluted models, clearly
distinct from that of the pure model, thus providing a complete
set of universality in disordered systems.
\end{abstract}

 \maketitle

Understanding the role of impurities on the nature of phase
transitions is of great importance, both from experimental and
theoretical perspectives. First-order phase transitions are known
to be significantly softened under the presence of quenched
randomness~\cite{aizenman-89,hui-89,chen-92,cardy-97,chatelain-98},
whereas continuous transitions may have their exponents altered
under random fields or random bonds~\cite{harris-74,chayes-86}.
There are some very useful phenomenological arguments and some,
perturbative in nature, theoretical results, pertaining to the
occurrence and nature of phase transitions under the presence of
quenched randomness~\cite{hui-89,dotsenko-95,jacobsen-98}.
Historically, the most celebrated criterion is that suggested by
Harris~\cite{harris-74}. This criterion relates directly the
persistence, under random bonds, of the non random behavior to the
specific heat exponent $\alpha_{p}$ of the pure system. According
to this criterion, if $\alpha_{p}>0$, then disorder will be
relevant, i.e., under the effect of the disorder, the system will
reach a new critical behavior. Otherwise, if $\alpha_{p}<0$,
disorder is irrelevant and the critical behavior will not change.
Pure systems with a zero specific heat exponent ($\alpha_{p}=0$)
are marginal cases of the Harris criterion and their study, upon
the introduction of disorder, has been of particular
interest~\cite{MK-99}. The paradigmatic model of the marginal case
is, of course, the general random two-dimensional (2D) Ising model
and this model has been extensively debated~\cite{gordillo-09}.

Respectively, the 3D Ising model with quenched randomness - which
is a clear case in terms of the Harris criterion having a positive
specific heat exponent in its pure version - has also been
extensively studied using Monte Carlo (MC)
simulations~\cite{landau-80,chow-86,heuer-90,hennecke-93,ball-98,wiseman-98,calabrese-03,berche-04}
and field theoretical renormalization group
approaches~\cite{folk-00,pakhnin-00,pelissetto-00}. Especially,
the diluted model can be treated in the low-dilution regime by
analytical perturbative renormalization group
methods~\cite{newman-82,jug-83,mayer-89}, where a new fixed point
independent of the dilution has been found, yet for the strong
dilution regime only MC results remain valid. Historically, the
first numerical studies of the model suggested a continuous
variation of the critical exponents along the critical line, but
it became clear, after the work of Ref.~\cite{heuer-90}, that the
concentration-dependent critical exponents found in MC simulations
are the effective ones characterizing the approach to the
asymptotic regime. Note, here, that a crucial problem of the new
critical exponents obtained in these studies is that the ratios
$\beta/\nu$ and $\gamma/\nu$ occurring in finite-size scaling
(FSS) analysis are almost identical for the disordered and pure
models. In fact, for the pure 3D Ising model, accurate values
are~\cite{guida-98}: $\nu=0.6304(13)$, $\beta/\nu=0.517(3)$,
$\gamma/\nu=1.966(3)$, and $\alpha=0.1103(1)$. Respectively, for
the site- and bond-diluted model, the most accurate sets of
asymptotic exponents $(\beta/\nu,\gamma/\nu,\nu)$ have been given
by the extensive numerical works of Ballesteros \emph{et
al.}~\cite{ball-98} and Berche \emph{et al.}~\cite{berche-04} are:
$(0.519(3),1.963(5),0.6837(53))$ and $(0.515(5),1.97(2),0.68(2))$.

The above estimates of critical exponents provided evidence that
the 3D Ising model with quenched uncorrelated disorder belongs to
a \emph{single universality class}, distinct from that of the pure
model, as also indicated by the Harris criterion, independent of
the considered disorder distribution. Yet, in a very recent paper
Murtazaev and Babaev~\cite{babaev-09} using MC simulations and FSS
methods on the site-diluted model the above view was contradicted
and these authors suggested that the model has two regimes of
critical behavior universality, depending on the nonmagnetic
impurity concentration. Motivated by the above contradictions and
the great theoretical interest of the existence of universality
classes in disordered models, we have chosen to investigate the 3D
Ising model with bond disorder in order to compare all these three
kinds of disorder (site-, bond-dilution and bond disorder) and to
verify whether these lead to the same set of new critical
exponents, as would be, in principle, expected by universality
arguments~\cite{berche-04}. In this contribution we show that,
indeed, the 3D Ising model with quenched, uncorrelated bond
disorder belongs to the same universality class as the site- and
bond-diluted models, defining in this way a complete universality
class in disordered spin models.

In the following we consider the 3D bond-disorder Ising model
whose Hamiltonian with uncorrelated quenched random interactions
can be written as
\begin{equation}
\label{eq:1} \mathcal{H}=-\sum_{\langle ij
\rangle}J_{ij}s_{i}s_{j},
\end{equation}
where the spin variables $s_{i}$ take on the values $-1,+1$,
$\langle ij \rangle$ indicates summation over all nearest-neighbor
pairs of sites, and the ferromagnetic interactions $J_{ij}>0$
follow a bimodal distribution of the form
\begin{equation}
\label{eq:2}
\mathcal{P}(J_{ij})=\frac{1}{2}~[\delta(J_{ij}-J_{1})+\delta(J_{ij}-J_{2})],
\end{equation}
where $J_{1}+J_{2}=2$, $J_{1}>J_{2}$, and $r=J_{2}/J_{1}$ reflects
the strength of the bond randomness. Additionally, we fix
$2k_{B}/(J_{1}+J_{2})=1$ to set the temperature scale. The value
of the disorder strength considered throughout this work is
$r=1/3$.

Resorting to large scale MC simulations is often necessary,
especially for the study of the critical behavior of disordered
systems. It is also well known that for such complex systems
traditional methods become inefficient and thus in the last few
years several sophisticated algorithms, some of them are based on
entropic iterative schemes, have been proven to be very
effective~\cite{newman-99}. The present numerical study has been
carried out by applying our recent and efficient entropic
scheme~\cite{fytas-08a}. In this approach we follow a two-stage
strategy of a restricted entropic sampling~\cite{malakis-04} based
on the Wang-Landau (WL) algorithm~\cite{wang-01}. As we do not
wish to reproduce here the details of our implementation, we give
only a brief discussion on the nature of the WL method. The usual
WL recursion proceeds by modifying the density of states $G(E)$
according to the rule $G(E)\rightarrow f G(E)$ and initially one
chooses $G(E)=1$ and $f=f_{0}=e$. Once the accumulative energy
histogram is sufficiently flat, the modification factor $f$ is
redefined as: $f_{j+1}=\sqrt{f_{j}}$, with $j=0,1,\ldots$ and the
energy histogram reset to zero until $f$ is very close to unity
(i.e. $f=e^{10^{-8}}\approx 1.000\; 000\; 01$). Once $f$ is close
enough to unity, systematic deviations become negligible. However,
the WL recursion violates the detailed balance from the early
stages of the process and care is necessary in setting up a proper
protocol of the recursion. In spite of the fact that the WL method
has produced very accurate results in several models, it is fair
to say that there is not a safe way to access possible systematic
deviations in the general case. This has been pointed out and
critiqued in a recent review by Janke~\cite{janke-08}. However,
from our experience and especially from our recent studies on 2D
and 3D disordered spin models~\cite{fytas-10}, the WL
implementation followed in these papers has produced excellent
results, enabling the discrimination between competing theoretical
predictions on that model.
\begin{figure}[htbp]
\includegraphics*[width=9 cm]{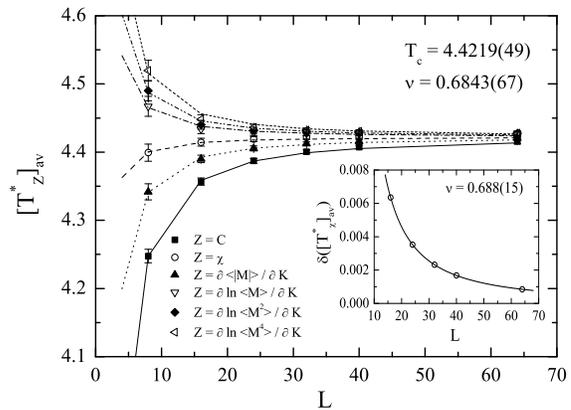}
\caption{\label{fig:1} Shift behavior of several pseudocritical
temperatures as defined in the text. The error bars reflect the
sample-to-sample fluctuations. The inset shows the FSS of the
sample-to-sample fluctuations of the pseudocritical temperature of
the magnetic susceptibility.}
\end{figure}
Using this combined approach we performed extensive simulations
for several lattice sizes $L\in\{8,16,24,32,40,64\}$, over large
ensembles of random realizations of the order of $1000-3000$. Each
disorder realization was simulated at least $10 - 20$ times with
different initial conditions to improve accuracy.

It is well known that, extensive disorder averaging is necessary
when studying random systems, where usually broad distributions
are expected leading to a strong violation of
self-averaging~\cite{wiseman-98,aharony-96}. A measure from the
scaling theory of disordered systems, whose limiting behavior is
directly related to the issue of self-averaging~\cite{wiseman-98}
may be defined with the help of the relative variance of the
sample-to-sample fluctuations of any relevant singular extensive
thermodynamic property $Z$ as follows:
$R_{Z}=([Z^{2}]_{av}-[Z]^{2}_{av})/[Z]^{2}_{av}$. Closely related
to the above issue of self-averaging in disordered systems is the
manner of averaging over the disorder. This non-trivial manner may
be performed in two distinct ways when identifying the finite-size
anomalies. The first way corresponds to the average over disorder
realizations ($[\ldots]_{av}$) and then taking the maxima
($[\ldots]^{\ast}_{av}$), or taking the maxima in each individual
realization first, and then taking the average
($[\ldots^{\ast}]_{av}$). Closing this brief outline, let us
comment on the statistical errors of our numerical data. The
statistical errors of our WL scheme on the observed average
behavior were found to be of small magnitude (of the order of the
symbol sizes) and thus are neglected in the figures. On the other
hand for the case $[\ldots^{\ast}]_{av}$ the error bars shown
reflect the sample-to-sample fluctuations.

We briefly present here our numerical results for the 3D
random-bond Ising model. Figure~\ref{fig:1} illustrates in the
main panel the shift behavior of $6$ pseudocritical temperatures
estimated via the second way of averaging discussed above, i.e. by
taking the average over the individual pseudocritical
temperatures. The error bars reflect the sample-to-sample
fluctuations.
\begin{figure}[htbp]
\includegraphics*[width=9 cm]{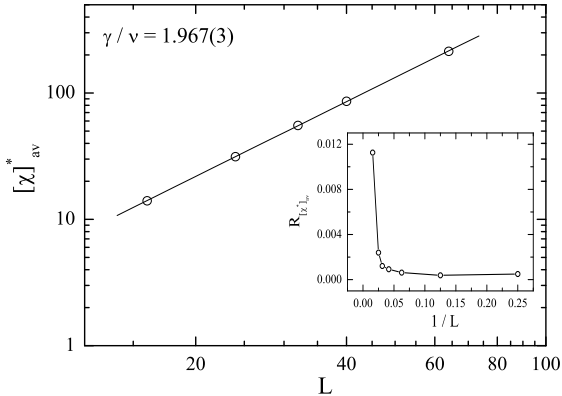}
\caption{\label{fig:2} FSS of the maxima of the disorder-averaged
magnetic susceptibility in a log-log scale. The inset shows the
limiting behavior of the ratio $R_{[\chi^{\ast}]_{av}}$.}
\end{figure}
The pseudocritical temperatures considered correspond to the peaks
of the following six quantities: specific heat $C$, magnetic
susceptibility $\chi$, derivative of the absolute order parameter
with respect to inverse temperature $K=1/T$: $\partial \langle
|M|\rangle/\partial K=\langle |M|H\rangle-\langle
|M|\rangle\langle H\rangle$~\cite{ferrenberg-91} and logarithmic
derivatives of the first ($n=1$), second ($n=2$), and fourth
($n=4$) powers of the order parameter with respect to inverse
temperature $\partial \ln \langle M^{n}\rangle/\partial K=\langle
M^{n}H\rangle/\langle M^{n}\rangle-\langle
H\rangle$~\cite{ferrenberg-91}. Fitting our data for the whole
lattice range to the expected power-law behavior
$[T^{\ast}_{Z}]_{av}=T_{c}+bL^{-1/\nu}$, where $Z$ stands for the
different thermodynamic quantities mentioned above, we find the
critical temperature to be $T_{c}=4.4219(49)$ well below the
critical temperature of the pure model and the estimate
$\nu=0.6843(67)$ for the critical exponent of the correlation
length. This value is in excellent agreement with the values
$0.6837(53)$ and $0.68(2)$ given by Ballesteros \emph{et
al.}~\cite{ball-98} and Berche \emph{et al.}~\cite{berche-04}.

Using the above sample-to-sample fluctuations of the
pseudocritical temperatures and the theory of FSS in disordered
systems as introduced by Aharony and Harris~\cite{aharony-96} and
Wiseman and Domany~\cite{wiseman-98}, one may further examine the
nature of the fixed point controlling the critical behavior of the
disordered system. According to the theoretical
predictions~\cite{aharony-96,wiseman-98}, the pseudocritical
temperatures $T_{Z}^{\ast}$ of the disordered system are
distributed with a width $\delta [T_{Z}^{\ast}]_{av}$, that
scales, in the case of a new random fixed point, with the system
size as $\delta([T_{Z}^{\ast}]_{av})\sim L^{-1/\nu}$. In the inset
of Fig.~\ref{fig:1} we plot these sample-to-sample fluctuations of
the pseudocritical temperature of the magnetic susceptibility. The
solid line shows a very good power-law fitting giving the value
$0.688(15)$ for the exponent $\nu$, which is also in very good
agreement with the value $0.6843(67)$ obtained via the classical
shift behavior and the most accurate estimates in the
literature~\cite{ball-98,berche-04}.
\begin{figure}[htbp]
\includegraphics*[width=9 cm]{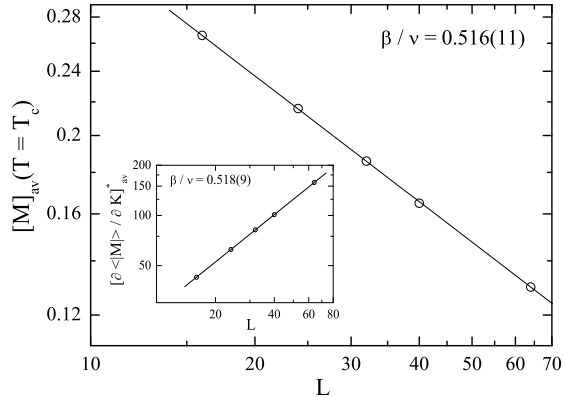}
\caption{\label{fig:3} FSS of the disorder-averaged order
parameter at the estimated critical temperature in a log-log
scale. The inset illustrates the FSS of the maxima of the
disorder-averaged inverse-temperature derivative of the absolute
order parameter.}
\end{figure}

In Figs.~\ref{fig:2} and \ref{fig:3} we provide estimates for the
magnetic exponent ratios of the model. In particular, in
Fig.~\ref{fig:2} we present the FSS of the maxima of the
disorder-averaged magnetic susceptibility $[\chi]_{av}^{\ast}$ in
a double logarithmic scale. The solid line presents a linear
fitting using the total lattice range spectrum, giving the
estimate $1.967(3)$ for the ratio $\gamma/\nu$
($[\chi]_{av}^{\ast} \sim L^{\gamma/\nu}$). The inset shows the
ratio $R_{Z}$, where $Z=[\chi^{\ast}]_{av}$ as a function of the
inverse lattice size indicating a strong violation of
self-averaging of the magnetic properties of the 3D Ising model
with bond disorder. Respectively, in Fig.~\ref{fig:3} we plot the
disorder-averaged magnetization at the estimated critical
temperature, as a function of the lattice size $L$, also in a
log-log scale. The solid line is a linear fitting
($[M]_{av}(T=T_{c}) \sim L^{-\beta/\nu}$) giving the value
$\beta/\nu=0.516(11)$. Additional estimate for the ratio
$\beta/\nu$ can be obtained from the FSS of the derivative of the
absolute order parameter with respect to inverse temperature which
is expected to scale as $L^{(1-\beta)/\nu}$ with the system
size~\cite{ferrenberg-91}. Thus, in the corresponding inset of
Fig.~\ref{fig:3} we plot the data for $\partial \langle
|M|\rangle/\partial K$ averaged over disorder as a function of
$L$, also in a double logarithmic scale. The solid line is a
linear fitting that gives an estimate of $0.518(9)$ for the ratio
$\beta/\nu$. Thus, overall the values for the ratios of the
magnetic exponents are in excellent agreement with the best
estimates for the site- ($\beta/\nu=0.519(3)$ and
$\gamma/\nu=1.963(5)$)~\cite{ball-98} and bond-diluted
($\beta/\nu=0.515(5)$ and $\gamma/\nu=1.97(2)$)~\cite{berche-04}
cases reinforcing the scenario of a single distinctive
universality in the 3D Ising model with quenched uncorrelated
disorder, independent of the disorder distribution.

\begin{figure}[htbp]
\includegraphics*[width=9 cm]{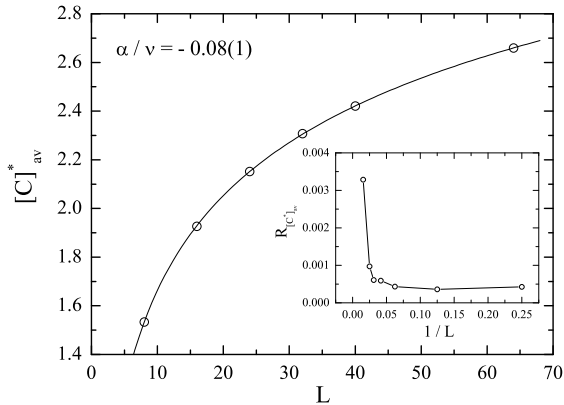}
\caption{\label{fig:4} FSS of the maxima of the disorder-averaged
specific heat. The inset shows the limiting behavior of the ratio
$R_{[C^{\ast}]_{av}}$.}
\end{figure}
Using our estimates for the critical exponents and the Rushbrooke
relation ($\alpha/\nu+2\beta/\nu+\gamma/\nu=2/\nu$) we estimate
the ratio $\alpha/\nu$ to be $-0.085(10)$. In Fig.~\ref{fig:4} we
present the FSS of the maxima of the disorder-averaged specific
heat data as a function of the linear size $L$. The solid line
shows a typical power-law fitting attempt of the form
$[C]_{av}^{\ast} \sim L^{\alpha/\nu}$, that gives a value
$\alpha/\nu=-0.08(1)$. This is a further reliability test of our
numerical method and extended simulations. The corresponding inset
of Fig.~\ref{fig:4} presents the ratio $R_{[C^{\ast}]_{av}}$ as a
function of the inverse linear size. As in the inset of
Fig.~\ref{fig:2}, also here a violation of self-averaging is
observed, yet this is, in absolute numbers, much smaller than that
of the magnetic susceptibility.

Summarizing, we have presented in this brief report concrete
evidence that the critical behavior of the 3D Ising model with
quenched uncorrelated disorder is controlled by a new random fixed
point, independent of the way randomness is implemented in the
system. This result has been obtained through extensive numerical
simulations and classical finite-size scaling techniques.
Particular interest was paid to the sample-to-sample fluctuations
of the pseudocritical temperatures model and their scaling
behavior that was used as a successful alternative approach to
criticality that verified the above scenario. Although we
acknowledge that Ballesteros \emph{et al.}~\cite{ball-98} and
Berche \emph{et al.}~\cite{berche-04} were the first to support
numerically the above view, we believe that the present
contribution puts a further significant step on this intensively
debated issue of universality in disordered spin models. Very
interesting would be also to study more complicated models, where
disorder couples to the order parameter, and one such prominent
candidate is the 3D random-field Ising model~\cite{vink-10}. For
these type of models, the existence of universality classes has
been severely questioned~\cite{sourlas-99}.

{}
\end{document}